\documentclass[fleqn,10pt]{wlscirep}
\usepackage[utf8]{inputenc}
\usepackage[T1]{fontenc}

\usepackage{caption}
\usepackage{subcaption}
\usepackage{multirow}

\usepackage{bbold}
\title{X-ray Image Generation as a Method of Performance Prediction for Real-Time Inspection: a Case Study}

\author[1,*]{Vladyslav Andriiashen}
\author[1,2]{Robert van Liere}
\author[1,3]{Tristan van Leeuwen}
\author[1,4]{K. Joost Batenburg}
\affil[1]{Computational Imaging, Centrum Wiskunde \& Informatica, Science Park 123, 1098 XG, Amsterdam, The Netherlands}
\affil[2]{Faculteit Wiskunde en Informatica, Technical University Eindhoven, Groene Loper 5, 5612 AZ Eindhoven, The Netherlands}
\affil[3]{Mathematical Institute, Utrecht University, Budapestlaan 6, 3584 CD Utrecht, The Netherlands}
\affil[4]{Leiden Institute of Advanced Computer Science, Leiden University, Niels Bohrweg 1, 2333 CA, Leiden, The Netherlands}

\affil[*]{vladyslav.andriiashen@cwi.nl}


\begin{abstract}
X-ray imaging can be efficiently used for high-throughput in-line inspection of industrial products. However, designing a system that satisfies industrial requirements and achieves high accuracy is a challenging problem. The effect of many system settings is application-specific and difficult to predict in advance. Consequently, the system is often configured using empirical rules and visual observations. The performance of the resulting system is characterized by extensive experimental testing. We propose to use computational methods to substitute real measurements with generated images corresponding to the same experimental settings. With this approach, it is possible to observe the influence of experimental settings on a large amount of data and to make a prediction of the system performance faster than with conventional methods. We argue that a high accuracy of the image generator may be unnecessary for an accurate performance prediction. We propose a quantitative methodology to characterize the quality of the generation model using POD curves. The proposed approach can be adapted to various applications and we demonstrate it on the poultry inspection problem. We show how a calibrated image generation model can be used to quantitatively evaluate the effect of the X-ray exposure time on the performance of the inspection system.
\end{abstract}
\begin{document}

\flushbottom
\maketitle

\thispagestyle{empty}

\section*{Introduction}

X-ray imaging systems are widely used for a high-throughput in-line inspection of industrial products on a conveyor belt\cite{mathanker2013x, mathiassen2011trends}. Unlike visible light cameras, X-ray radiation is able to penetrate the product and provide information about its internal structure. This is particularly useful in a foreign object detection (FOD) problem, where the goal is to inspect a \textit{main} object and determine whether it contains an undesirable region or a separate body - a \textit{foreign} object. Examples of foreign objects (FOs) are browning in apples, infestation in grain, and bones in poultry\cite{du2019x}. While X-ray inspection could achieve high accuracy in laboratory studies\cite{olakanmi2023applications, naresh2020use}, an industrial environment introduces many limitations, such as high-throughput, acceptable dose level, and physical space constraints. Designing a practically viable X-ray system - configuring the experimental equipment and the image analysis algorithm - is a challenging problem of balancing inspection accuracy and application constraints. Furthermore, this problem is highly application-specific; methods developed for one product are usually not effective for other products\cite{mathanker2013x}.

A standard approach to analyzing the inspection accuracy is extensive experimental testing. To achieve automated inspection without frequent manual checks, the inspection system has to be reliable. In the context of FOD, reliable systems should detect FOs with a low failure rate for a wide variety of products. The shape of the foreign object is not known in advance, and it could be located in different parts of the main object. The direction from which the product would be viewed is also not known in advance. All these variations need to be tested to confirm that the system is reliable. However, this time-consuming test would only be valid for a specific configuration of experimental equipment and image analysis. Any change in the hardware settings would have an unpredictable effect on the acquired images and the performance of the system. Designing an inspection system from scratch involves a large number of independent settings. Performing an experimental test for all possible permutations of settings is infeasible. A different approach is therefore required.

There is a rising interest in creating virtual representations of physical products and systems - digital twins\cite{jones2020characterising}. Digital twins of X-ray systems\cite{baldo2020digital, ahmed2022integrating, bircher2021high} are seen as a tool for predicting the X-ray image under the desired experimental settings without an experiment. Such a model could solve the system design problem. If the data could be generated significantly faster than the experimental acquisition time, then it would be practically possible to test a large number of different system settings. The problem of generating X-ray images with computational methods has been studied in detail with vastly different approaches\cite{bellon2007artist, gong2018rapid, bergback2013mcxtrace}. While it is possible to implement a highly accurate virtual model of an X-ray setup\cite{jan2011gate}, the computational cost is significant. As a result, sufficiently fast data generation would be too expensive and infeasible. This computational challenge leads to the question of how accurate the X-ray image generator needs to be to be used in practice. 

In some cases, approximately accurate generated images could even be used to substitute the real ones\cite{van2022inline, andriiashen2023ct}. It is possible that perfect correspondence is not necessary, as long as the features used by the image analysis algorithm are generated accurately. However, to conclude that the performance of the inspection system is the same (or similar) for two different datasets, a quantitative way of characterizing the performance is required. This problem is well-known in nondestructive testing, and Probability of Detection (POD) curves\cite{georgiou2007pod} are often used as a solution. POD curves provide a quantitative estimate of how the detection rate depends on the value of the product properties, and for which values reliable detection is expected. Product properties associated with reliable detection can be measured and compared to conclude if the performance on two datasets is the same.

We propose a novel data-driven method for efficient performance estimation of X-ray imaging systems under different experimental settings. Our contributions are as follows: We combine the theoretical foundations of X-ray image formation and X-ray data generation techniques to create images corresponding to specific experimental settings of the given inspection system. We show which properties of the system should be measured during calibration to make such generation possible. As an example of the FOD problem, we focus on the detection of small rib bones in chicken fillets with different X-ray exposure times. For the same experimental settings, a real dataset is acquired with the experimental setup; and a generated dataset is created with a calibrated generator and the dataset of high-quality real images. We show that the use of generated data yields quantitatively comparable performance estimates to the real data, and explore the performance for settings that were not tested. We discuss the advantages of the proposed approach over the standard experimental testing and how our method could be applied to other industrial problems.

\section*{Related work}

The problem of X-ray data generation is well-studied and can be solved with a variety of methods that are commonly categorized into probabilistic and deterministic. Probabilistic (Monte-Carlo) methods\cite{jan2011gate, bergback2013mcxtrace} require a significant amount of computational resources and extensive knowledge of the X-ray system and the studied object. Deterministic (ray-tracing) methods\cite{gong2018rapid, bellon2007artist} are faster by orders of magnitude but do not include many experimental effects present in a real X-ray acquisition. Furthermore, deterministic algorithms produce a noiseless image, and the noise pattern has to be generated separately. Properties of noise in X-ray images were studied in detail to estimate measurement uncertainty\cite{rodriguez2020review} and improve analysis of low-dose X-ray data\cite{ma2012variance}. Equations connecting mean value and variance in X-ray data are derived in literature and experimentally verified\cite{ma2012variance}.

Image analysis for the FOD is a sophisticated problem that can be addressed by many methods based on contrast\cite{song2004automated}, edge detection\cite{zohora2017circular}, and machine learning trained on manually extracted features. A large variety of main and foreign objects has to be incorporated into the algorithm to achieve the necessary robustness. Deep learning methods have been successfully applied to many computer vision tasks: first, Convolutional Neural Networks (CNN)\cite{tulbure2022review}, and more recently, Vision Transformers (ViT)\cite{dosovitskiy2020image}. Deep learning algorithms require a large amount of training data to determine the image features that are crucial to solving the task in question. With sufficient data coverage, these methods provide state-of-the-art results and achieve high accuracy, execution speed, and robustness.

A significant disadvantage of deep learning methods is a lack of interpretability. Some techniques, such as class activation maps, might correlate the algorithm's decision with some properties of the input image. In contrast, in conventional algorithms, such as edge localization with the Canny filter, the accuracy can be computed analytically based on the noise model\cite{kakarala1992achievable}. However, this is impossible for real problems with complex image morphology. In practice, a deep learning method has to be analyzed as a black box, and the performance analysis may benefit from statistical analysis used to assess the accuracy of human experts (e.g. ROC curves and POD curves\cite{georgiou2007pod}). Such methods imply that a sufficient variety of test cases can be provided to accurately estimate the performance of the algorithm.

Optimization of X-ray imaging has been studied in medical imaging to determine the geometry of experimental setup\cite{siewerdsen2000optimization} and dose of radiation\cite{gislason2010dose}. The existing work proposes different theoretical methods that explore the connection between the properties of the experimental setup and image quality. Image quality is usually defined by performing phantom measurements and computing different image features, such as the signal-to-noise ratio. However, the correlation between such metrics and the performance on real problems is rarely studied. A possible solution to this problem was presented in \cite{gupta2019predicting} where a binary classifier was trained to predict if an object would be correctly identified by a human expert based on a set of image quality metrics.

\section*{General concepts}

\begin{figure}[ht]
\centering
\includegraphics[width=0.6\linewidth]{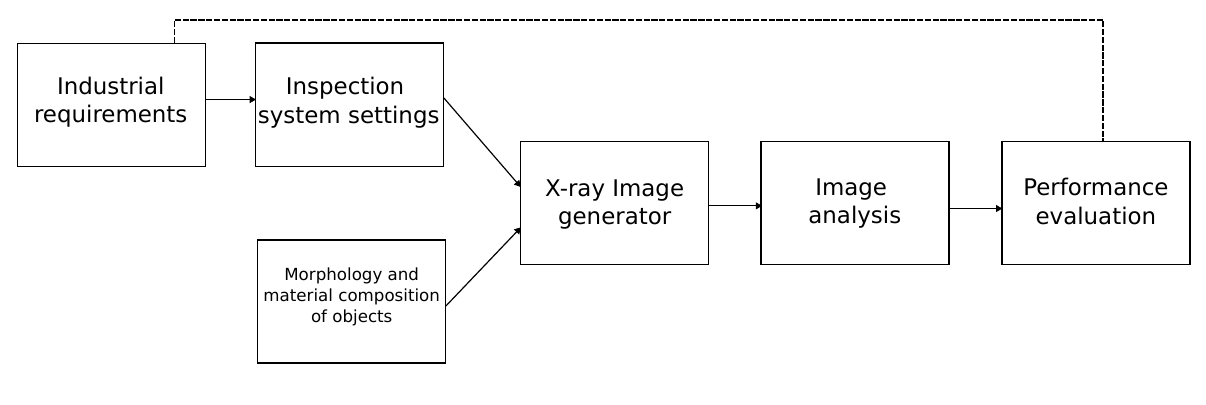}
\caption{ Overview of our approach for the FOD. The problem is defined by the physical properties of the objects to be analyzed and the inspection system settings. This leads to a variety of input images that have to be analyzed by the same image analysis algorithm. The performance of the algorithm is then evaluated as a function of the system design and the properties of the objects.}
\label{fig:method}
\end{figure}

Our approach to analyzing industrial FOD problems is illustrated on Fig. \ref{fig:method}. An applicable system has to satisfy industrial requirements. In the case of X-ray, we highlight a list of frequently relevant constraints:

\begin{itemize}
    \item Geometric - the inspection system should fit into the existing conveyor belt setup.
    \item Deposited dose - inspected objects should not absorb too much X-ray radiation.
    \item Radiation safety - properties of the X-ray source, such as voltage and power, are limited by the permits and radiational shielding to prevent interference with other parts of the factory.
    \item Resolution - the system is supposed to detect features of a certain size.
    \item Throughput - the inspection system should operate at a speed comparable to the conveyor belt.
\end{itemize}

These requirements directly affect the settings of the X-ray system: possible magnification, X-ray source power, exposure time etc. The same requirements can be satisfied in many ways leading to many possible systems. From the industrial point of view, the FOD problem is formulated by defining the physical properties of the objects on the conveyor belt. For X-ray, the important ones are material composition (what is the object made of) and morphology (what is the shape of the object). 

System settings and object properties influence the resulting X-ray image acquired during the inspection procedure. We argue that this image could be computed with a properly configured image generator without performing an experiment. The decision about the FO presence is made by the image analysis algorithm that takes the X-ray image as input. Image analysis is only indirectly influenced by the system settings and object properties. Instead, the accuracy of the analysis is mainly determined by image features. Finally, there is a performance evaluation step that involves testing the inspection system on a large variety of products with and without FOs. The performance evaluation concludes whether the proposed system design is applicable to the problem taking into account the industrial requirements and the desired accuracy level.

In the context of Fig. \ref{fig:method}, we consider the image generator to be sufficiently accurate when it yields the same (or quantitatively comparable) performance estimate as the real inspection system with the same settings and objects. To make such a generator, we start with a model describing an X-ray image formation. This model contains system-specific parameters that we propose to extract from the calibration of the real system. The goal of the calibration is to connect an abstract model of X-ray interactions to the particular X-ray imaging device under different system settings. When a sufficiently accurate generator is available, the industrial requirements should be translated into different sets of feasible system settings. For each set of settings, a performance estimate is computed using the image generator. If the requirements constrained the system too much, all estimates would not be sufficiently accurate, meaning that the problem has no practical solution under the current limitations. Otherwise, we could use the settings that lead to the best performance and confirm that they work in the real system.

\section*{Methods}

\begin{figure}[ht]
\centering
\includegraphics[width=0.6\linewidth]{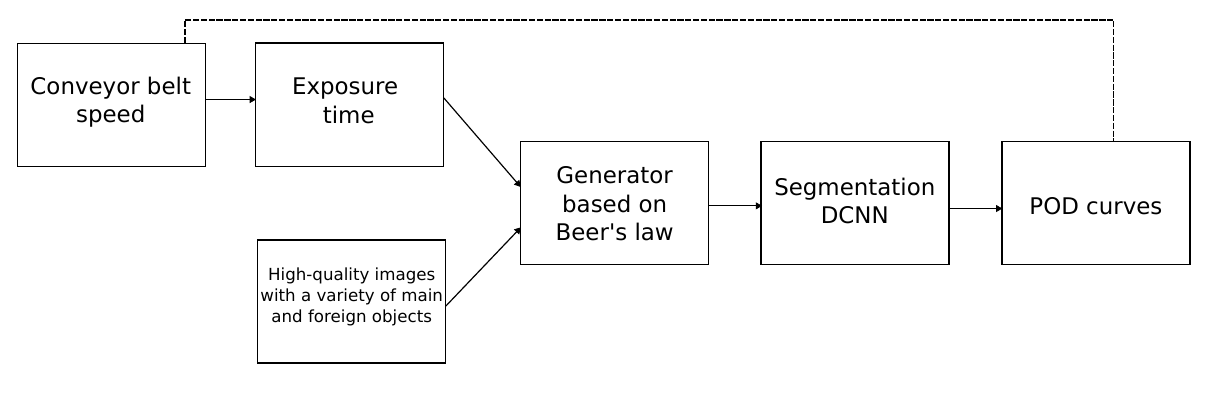}
\caption{System design problem for the detection of bone fragments in chicken fillets under different levels of throughput. This figure illustrates the technical choices made to solve this problem: high-quality X-ray images as a representation of object morphology, an image generator based on Beer's law, DCNN as the image analysis method, and a POD curve as a performance metric.}
\label{fig:case_study}
\end{figure}

To illustrate the proposed approach of predicting inspection performance using image generation, we concentrate on a specific case study: detecting bone fragments in chicken fillets under different conveyor throughput levels. The application-specific version of Fig. \ref{fig:method} is shown on Fig. \ref{fig:case_study}. We consider a single industrial requirement - throughput of the system - that influences the exposure time possible for an X-ray image acquisition. A large number of high-quality X-ray images is used to represent a variety of fillets and bones that could be found on a conveyor belt. We consider a dual-energy X-ray setup as the inspection system. An image generator based on Beer's law with the mixed Poisson-Gaussian noise model is proposed to create images for different values of exposure time. The image analysis is performed by a segmentation DCNN, and the detection performance is analyzed with POD curves.

\subsection*{Industrial requirements and inspected objects}

For in-line product inspection, the X-ray imaging system is built around the high-throughput conveyor belt. The imaging system consists of an X-ray tube which acts as a source of X-ray radiation and a detector which acquires an image by interacting with X-ray passing through the object. The conveyor belt speed is one of the most important application properties. The processing line is expected to run with a certain number of products per unit of time, and the inspection system must operate under this constraint. For the pixel size $a$ and the conveyor belt speed $v$, the measurement time $t$ is limited according to the equation:
\begin{equation}
\label{exp_time}
    t \ll \frac{a}{v}.
\end{equation}
For line detectors that capture an image row by row, a longer exposure time would lead to an incorrect image that misses parts of the object. Furthermore, if the exposure time is close to the upper limit of $\frac{a}{v}$, there will be significant motion blur in the image.

Similarly to the FOD in other agricultural products, poultry inspection does not have a rigorous definition of main and foreign objects. Chicken fillets are similar in shape to a human expert, but the exact shape is unknown prior to the inspection. The chemical composition of chicken meat is approximately known. Bone fragments could be found in a variety of locations inside the fillet depending on the cutting process. We choose a data-driven way to describe the variety of main and foreign objects, and their relative positions. A reference dataset is used as a problem definition. This dataset contains high-quality X-ray images of different chicken fillets with different bone fragments in various positions. The definition of high quality will be given later, and it refers to the low noise level to capture the X-ray attenuation properties of fillets and bones. The goal of the image generator is to transform a reference high-quality image into the image corresponding to a particular value of the exposure time.

\subsection*{Image formation}

To estimate the influence of the imaging system and object properties, we use a model of X-ray imaging based on Beer's law. This model is commonly used in computed tomography. It assumes that the X-ray intensity $I$ decays after passing through the object according to Beer's law
\begin{equation}
\label{eq:beers_law}
    I = I_0 \exp{\left( -\int_l \mu(x) dx \right)},
\end{equation}

\noindent where $I_0$ is the intensity of the incident radiation, $l$ is the X-ray trajectory, and $\mu(x)$ is the distribution of the attenuation coefficient. The goal of imaging is to estimate the properties of the object based on $\mu(x)$, which is connected to the measurable value of $I$.

For further analysis, we will make a number of assumptions about the studied object. $\mu(x)$ is a combination of attenuation in the main and the foreign object. We assume that both are homogeneous. A homogeneous object consists of a single material with a constant attenuation rate $\mu$ and occupies a volume $V$. Thus, if the studied object has no FO and consists of only one material,
\begin{equation}
    \mu(x) = \mu \mathbb{1}_V (x), \text{ where } \mathbb{1}_V (x) = \begin{cases}
    1 & \text{if $x \in V$} \\
    0 & \text{otherwise}
\end{cases}.
\end{equation}
While $V$ is the physical shape of the object, which remains the same regardless of the imaging method, $\mu$ depends on both the object and image acquisition. X-ray tubes produce X-ray radiation with a wide range of energies characterized by the energy spectrum $I_0 \Phi(E)$. For any material with a known chemical formula, an attenuation curve $\mu(E)$ can be computed to estimate the attenuation probability for every X-ray energy. It can be shown\cite{alles2007beam} that $\mu$ in Beer's law is approximately defined by the equation
\begin{equation}
    \label{eq:poly_xray}
    \mu = \int_E g(E) D(E) \Phi(E) \mu(E) dE
\end{equation}

\noindent that also includes the detector properties: gain $g(E)$ and sensitivity $D(E)$.

The intensity $I$ in Eq. \ref{eq:beers_law} represents the true value, but it is not necessarily equal to the value measured by the imaging system. The generation of X-ray photons $I_0$ follows Poisson distribution. It can be shown\cite{whiting2002signal, whiting2006properties} that the number of registered photons $I$ also follows Poisson distribution under certain conditions. Additionally, there is an electronic noise independent of the number of detected X-ray photons. Consequently, the measured signal $y$ follows a mixed Poisson-Gaussian distribution
\begin{equation}
\label{eq:noise_def}
    y = p + b \sim \frac{1}{\lambda} \mathcal{P}(\lambda I) + \mathcal{N}(d_e, \sigma_e),
\end{equation}
\noindent where $\lambda$ denotes the detector gain and converts the number of X-ray photons following the Poisson distribution to the measured intensity units, $d_e$ is often referred to as the darkfield signal (image intensity without X-ray irradiation), and $\sigma_e$ is a standard deviation of the darkfield signal. Furthermore, we add a convolution with a Gaussian kernel
\begin{equation}
\label{eq:noise_blur}
    y_{blur} = y \circledast h, \qquad h(x,y) = a e^{-\frac{x^2+y^2}{2\sigma^2}}
\end{equation}
\noindent to represent all effects influencing the Point Spread Function (e.g. focal spot of the tube and scintillator processes).

The image formation model defined by Eq. \ref{eq:beers_law} and \ref{eq:noise_def} splits the contribution of object properties ($V$ and $\mu(E)$) and imaging system design ($\Phi(E)$, $I_0$, and detector parameters $g(E)$, $D(E)$, $\lambda$, $d_e$, $\sigma_e$). While the former are constant and define the FOD problem, the latter can be optimized to provide the best image quality. All detector parameters in this model are inherent properties of the chosen equipment and can not be changed without replacing a detector. This type of optimization is beyond the scope of this paper because it is mostly defined by the price of the detector and the available budget. However, the incident beam intensity $I_0$ and the energy spectrum $\Phi(E)$ can be adjusted without changing the equipment.

The incident beam intensity $I_0$ depends on the geometry of the imaging system and the tube settings. X-ray tubes produce radiation uniformly distributed in a cone of height $d$ - the distance between the source and the pixel. The flux $j$ depends slightly on the voltage and is proportional to the tube current $i$. The X-ray radiation is collected in a square pixel of size $a$ over the exposure time $t$. Thus, the incident beam intensity $I_0$ is given by
\begin{equation}
\label{eq:flux_intensity}
    I_0 \approx j \frac{a^2}{4\ pi d^2} t \propto i t \frac{a^2}{d^2},
\end{equation}

\noindent if the pixel is close to the center, and cone beam artifacts can be ignored. This equation offers many ways to control the incoming intensity: changing the imaging resolution ($a$), the exposure time ($t$), the geometric configuration of the system ($d$), and the current $i$.

\subsection*{Dual-energy image segmentation}

If the inspected object consists of the main object with properties $(\mu_m, V_m)$ and the foreign object with properties $(\mu_f, V_f)$, Eq. \ref{eq:beers_law} transforms into 
\begin{equation}
    I = I_0 \exp{\left( -\mu_m \int_l \mathbb{1}_{V_m} (x) dx - \mu_f \int_l \mathbb{1}_{V_f} (x) dx \right)}.
\end{equation}

The image is usually analyzed after post-processing according to the equation
\begin{equation}
\label{eq:cor_intens}
    M = -\log \frac{I}{I_0} = \mu_m L_m + \mu_f L_f,
\end{equation}

\noindent where $L = \int_l \mathbb{1}_{V_m} (x) dx$ is the thickness of the object $V$ along the ray $l$. 

In dual-energy acquisition, two images are acquired with different energy spectra $\Phi(E)$. This leads to different flatfield values $I_0^a$ and $I_0^b$ and attenuation coefficients $\mu^a$ and $\mu^b$. While $M$ is proportional to the object thickness, the quotient $R$ of corrected images is constant for homogeneous materials

\begin{equation}
\label{eq:mono_quot}
    \begin{cases}
        R_m = \frac{M^a_m}{M^b_m} = \frac{\mu_m^a L_m}{\mu_m^b L_m} = \frac{\mu_m^a}{\mu_m^b} \\[6pt]
        R_f = \frac{\mu_f^a}{\mu_f^b}
    \end{cases}
\end{equation}

For the main object with the present foreign object, the quotient image changes to
\begin{equation}
    R = \frac{M^a}{M^b} = \frac{\alpha \beta R_f + R_m}{\alpha \beta + 1},
\end{equation}
where $\alpha = \frac{L_f}{L_m}$ and $\beta = \frac{\mu_f^b}{\mu_m^b}$. The presence of the foreign object can be detected by observing a difference between $R$ and $R_m$ given by the equation
\begin{equation}
\label{eq:dr_def}
    \Delta R = \frac{\alpha \beta (R_f - R_m)}{\alpha \beta + 1}.
\end{equation}

Following Eq. \ref{eq:dr_def}, the FO can be found in a quotient image by locating a connected cluster of pixels with non-zero $\Delta R$ surrounded by noise. The magnitude of the background noise can be computed by estimating noise in Eq. \ref{eq:noise_def} and propagating it through Eq. \ref{eq:cor_intens} and \ref{eq:dr_def}. However, there are many practical limitations hampering the viability of this approach for real data. Eq. \ref{eq:dr_def} is written for a single pixel. A foreign object covers a region in the image with different values of thickness (and $\alpha$) leading to a spatial distribution of $\Delta R$. It is implied by Eq. \ref{eq:poly_xray} that $\mu$ is constant, but it is not the case due to beam-hardening\cite{andriiashen2021unsupervised}. Finally, the quotient image has a high noise level, especially near the boundaries where both $M^a$ and $M^b$ are small.

We use DCNNs to solve the FOD problem. Given the images $M^a$ and $M^b$, the goal of the DCNN is to output the image mask of $L_f > 0$. To train the DCNN, we provide a large set of training data using the manual segmentation serving as ground-truth. Due to generalization from a large amount of data, DCNNs are able to achieve high segmentation accuracy despite all previously mentioned challenges. 

\subsection*{Performance analysis}

Due to the large number of trainable parameters and low interpretability of DCNNs, it is difficult to predict how accurate the output would be given the image features. Eq. \ref{eq:dr_def} gives an example of one feature $\Delta R$ that could be used to distinguish FO from the main object. Thus, we assume that the accuracy should be higher for large values of $\Delta R$ and lower for small $\Delta R$. As a result, there should be a value of $\Delta R$ for which the algorithm is expected to work with high accuracy. We use the Probability of Detection analysis to characterize the performance of DCNNs and find out when they reach high accuracy.

A segmentation DCNN outputs a segmented image where every pixel is marked as corresponding to either the main object, foreign object or the background. We convert the segmentation into a binary result - detection. If the image contains no FO, the segmentation should not have any pixels marked as the FO to be considered correct. If there is an FO, the correct segmentation should have FO pixels that do not completely miss the correct FO location (Recall $>$ 10\%).  

We assume that the Probability of Detection $P$ depends on $\Delta R$ according to the equation
\begin{equation}
    g(P) = c_0 + c_1 \Delta R,
\end{equation}
where $g(P) = \log(-\log(1-P))$ is an S-shaped link function. Parameters $c_0$ and $c_1$ are determined using the Maximum Likelihood Estimation based on a sequence of $\Delta R$ and the corresponding detection outcomes (success with probability $P$ and a failure with probability $1-P$). The value of $P$ describes the performance of the system. If the system design ensures that the value of $\Delta R$ is high enough for FOs of interest, then the value of $P$ will be large enough for consistent detection. Thus, the value of $\Delta R$ can be seen as image quality in the context of detection.

We use the value of $\Delta R$ corresponding to $P = 90\%$ as the performance metric. $\Delta R$ characterizes all components of the inspection system: both physical properties of the studied object and experimental settings. The main disadvantage of using $\Delta R$ as a performance metric lies in the fact that a large variety of foreign objects could have the same contrast value. As shown in Eq. \ref{eq:dr_def}, the contrast depends on the FO thickness ($L_f$ in $\alpha$), its location in the main object ($L_m$ in $\alpha$). Furthermore, there is an implicit effect of the FO shape since $\Delta R$ is distributed over the image region. This problem is not unique to our approach of performance evaluation and stems from the X-ray image formation.

POD curves provide a way to evaluate the accuracy of the image generator. Two real images of the same object are never the same due to noise fluctuations. Thus, a direct comparison between real and generated images would require to check whether the noise distribution is correct. However, we only need generated images to estimate the performance of DCNNs. Thus, if the POD curve based on real data is indistinguishable from one derived from generated data, the generator is sufficiently accurate.

\section*{Results}

\subsection*{System calibration}

All experiments were performed at the FleX-ray laboratory\cite{coban2020explorative} of Centrum Wiskunde en Informatica in Amsterdam, the Netherlands. Unlike line detectors commonly used in industrial applications, FleX-ray contained a planar detector. The area of the detector was 143~mm $\times$ 114~mm, the projection size was 956~px $\times$ 760~px with a resolution of 150~$\mu$m. We performed the detector calibration\cite{konstantinidis2012dexela} to determine parameters from Eq. \ref{eq:noise_def} characterizing the setup.

The calibration was performed with tube voltages of 40~kV and 90~kV with an additional filter of 50~$\mu$m copper. The Source-to-Object distance was 990~cm and the Source-to-Detector was 1059~cm leading to a small magnification of 1.07 and close to the maximum field of view. We acquired a series of 5000 flatfield images with the same experimental settings. The series was converted into a distribution of the mean intensity value $\overline{y}$ and its standard deviation $\sigma_y$ over the detector plane. We repeated the flatfield measurement for different values of the tube current leading to a sequence of $\{\overline{y}; \sigma_y\}$ for every pixel of the detector. Using Eq. \ref{eq:noise_def} and linear regression, we extracted the values of the noise parameters:

\begin{equation}
\label{eq:calibration}
    \begin{cases}
       \overline{y} = I + d_e \\
        \sigma_y^2 = g I + \sigma_e^2
    \end{cases} \Rightarrow
    \sigma_y^2 = g (\overline{y} - d_e) + \sigma_e^2.
\end{equation}

\begin{figure}[ht]
\centering
\includegraphics[width=\linewidth]{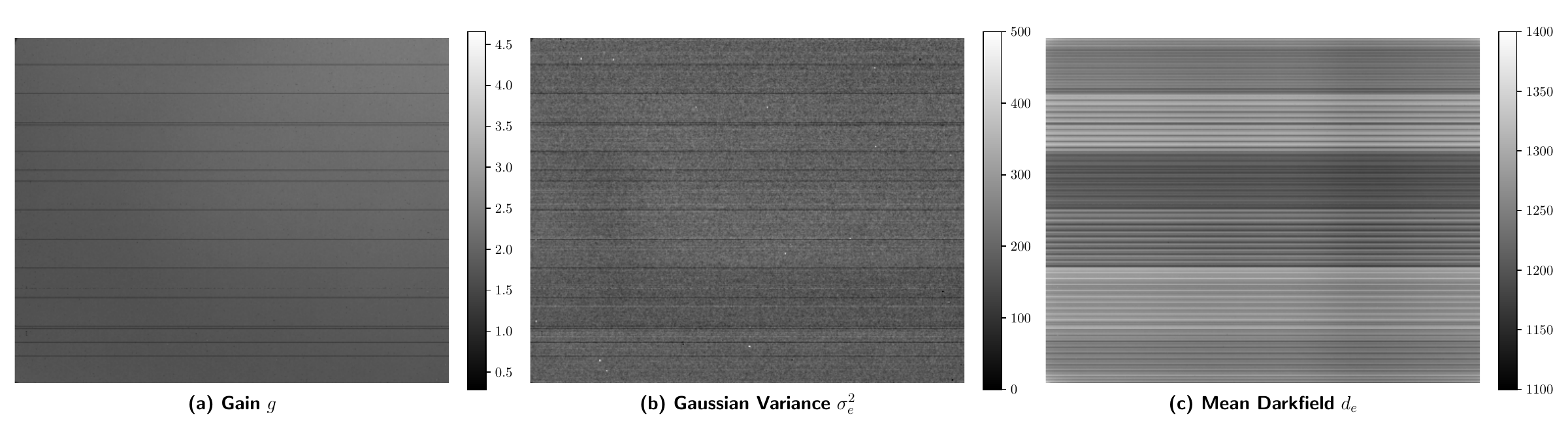}
\caption{Distributions of the extracted noise parameters over the detector plane: (a) gain $g$, (b) variance of darkfield $\sigma_e^2$, and (c) the mean values of darkfield $d_e$.}
\label{fig:noise_param}
\end{figure}

The distributions of the extracted parameters are shown on Fig. \ref{fig:noise_param}. These values are unique to our experimental setup, but the procedure can be applied to any detector. Furthermore, we computed an approximate value of the blur radius $\sigma = 0.8$ px (Eq. \ref{eq:noise_blur}). To obtain the estimate, we shifted a rectangular window around the image and computed the covariance matrix between the central pixel and the neighboring pixels. We made a fit of the covariance as a function of distance between pixels with a Gaussian function and used its RMS width as an approximation for $\sigma$.

Following Eq. \ref{eq:flux_intensity}, we used the flatfield measurement to determine the connection between the incident beam intensity $I_0$ and the exposure time $t$ in a form
\begin{equation}
\label{eq:beam_time}
    I_0 = k t.
\end{equation}
It was computed that $k = 0.58$ 1/ms for 40~KV and 40~W, and $k = 3.86$ 1/ms for 90~KV and 45~W.

\begin{figure}[ht]
\centering
\includegraphics[width=\linewidth]{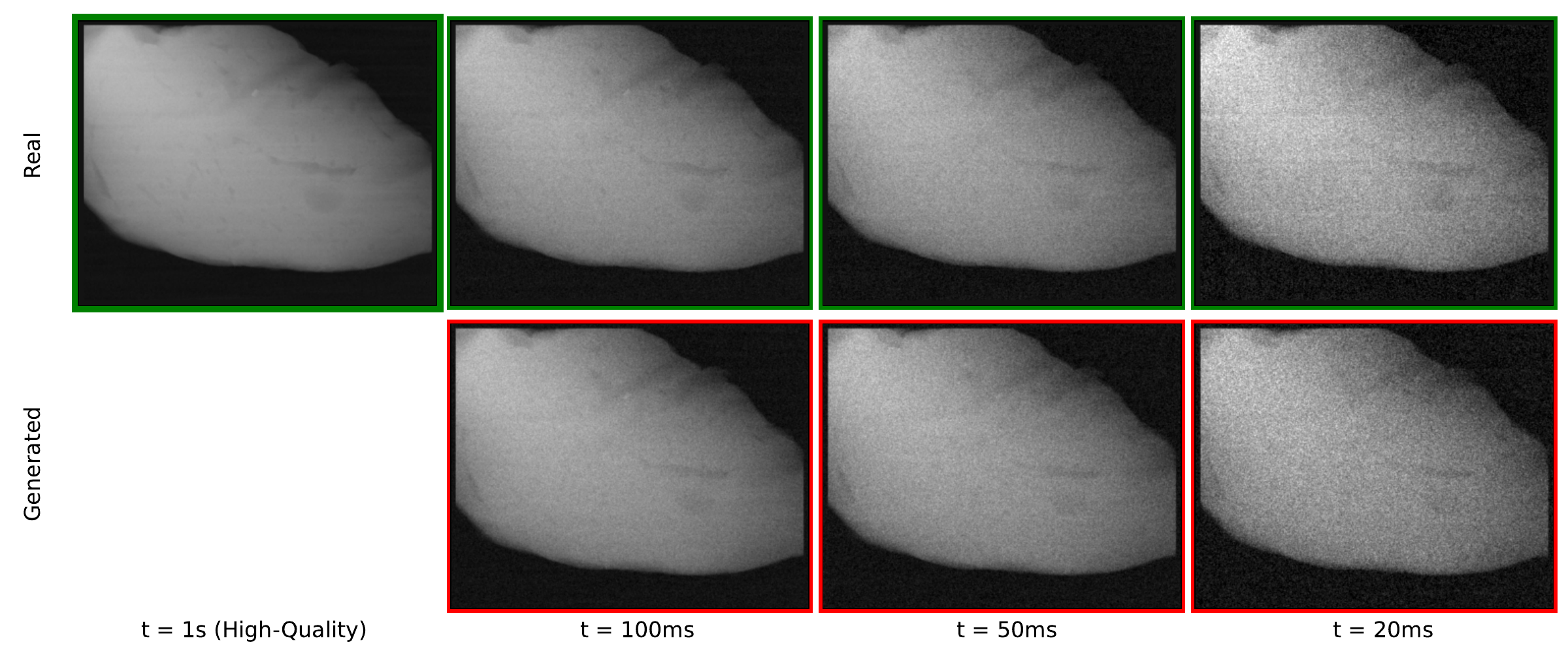}
\caption{Comparison of the generated and real images corresponding to different values of exposure time: 1s, 100ms, 50ms, 20ms. The inspected object is a chicken fillet contraining a bone fragment.}
\label{fig:comp_gen}
\end{figure}

After calibration, a noisy image corresponding to a certain exposure time $t$ could be produced using a reference high-quality image. We considered an image to be high-quality if the intensity values are large enough so that $\sigma_y \ll \overline{y}$, and the exposure time of 1~s was considered to be sufficient. To generate a noisy image for the exposure time $t$, we estimated the beam intensity $I_0$ using Eq. \ref{eq:beam_time}. Then according to Beer's law (Eq. \ref{eq:beers_law}) and the reference image we computed the mean estimated values of intensity $I$. Finally, the noise was added using a random number generator according to the distribution from Eq. \ref{eq:noise_def}. An example of the high-quality reference image ($t = 1$s) and noisy real and generated images for different values of exposure time (100ms, 50ms, 20ms) is shown on Fig. \ref{fig:comp_gen}. According to our observations, including the blurring step from Eq. \ref{eq:noise_blur} improved visual similarity between the real and the generated images.

\subsection*{Bone fragment detection in chicken fillets}

The image generation approach was tested on the problem of detecting bone fragments in chicken fillets. In this task, full chicken fillets were imaged with X-ray, and some of them contained pieces of rib bone of different lengths. The goal was to detect as small bones as possible to satisfy health safety guidelines. Furthermore, the inspection could not use a long exposure since it would limit the conveyor belt throughput.

The X-ray inspection in the industrial environment would commonly use a line detector below the conveyor belt and a tube above it, so a chicken fillet is parallel to the detector surface. We used a laboratory setup to acquire the data, and to achieve similar images we have attached fillets vertically to a plastic board. Pieces of rib bones were cut from a chicken carcass. Bone fragments were placed on top and inside the fillet manually and randomly.

We have acquired 2 datasets with a total of 338 X-ray projections of chicken fillets. The experimental geometry and tube properties were the same as during calibration. The goal of the first dataset was to capture a variety of different products and FOs to train a DCNN. Furthermore, the first dataset was used to measure the fraction of false positives (products without FO where an FO was incorrectly detected). The second dataset presented a narrower task where two small bone fragments were placed in different locations of two fillets. The purpose of this dataset was to find the threshold value of $\Delta R$ that guarantees reliable detection. 

The data acquisition was performed in two stages. First, we made a dataset consisting of 163 projections containing a bone fragment and 91 projections of boneless fillets. There were 14 different chicken fillets and 44 bone fragments of different sizes. All projections were recorded with an exposure time of 1~s. The size of the bone fragments ranged from 1.5~mm to 11~mm. The second dataset was made with two chicken fillets and two small bone fragments (2~mm and 3~mm) that were placed in different regions of the same fillet. This dataset contained 84 projections (two boneless and 82 with a bone fragment) acquired with 4 different values of exposure time: 1~s, 100~ms, 50~ms, and 20~ms.

\subsection*{Foreign object detection}

To solve the FOD problem with DCNNss, we have used Segmentation Models package\cite{Iakubovskii:2019}. DeepLabV3Plus with EfficientNet-B0 encoder and encoder depth of 5 (4.5~M trainable parameters) was chosen as the DCNN architecture. The network was trained on the first dataset with a split between training and validation data. The stopping criterion was achieving the minimum Dice's loss on the validation subset. To account for the random nature of the training process, the network was trained 100 times with the same data and different random seeds.

\begin{figure}[ht]
\centering
\includegraphics[width=0.95\linewidth]{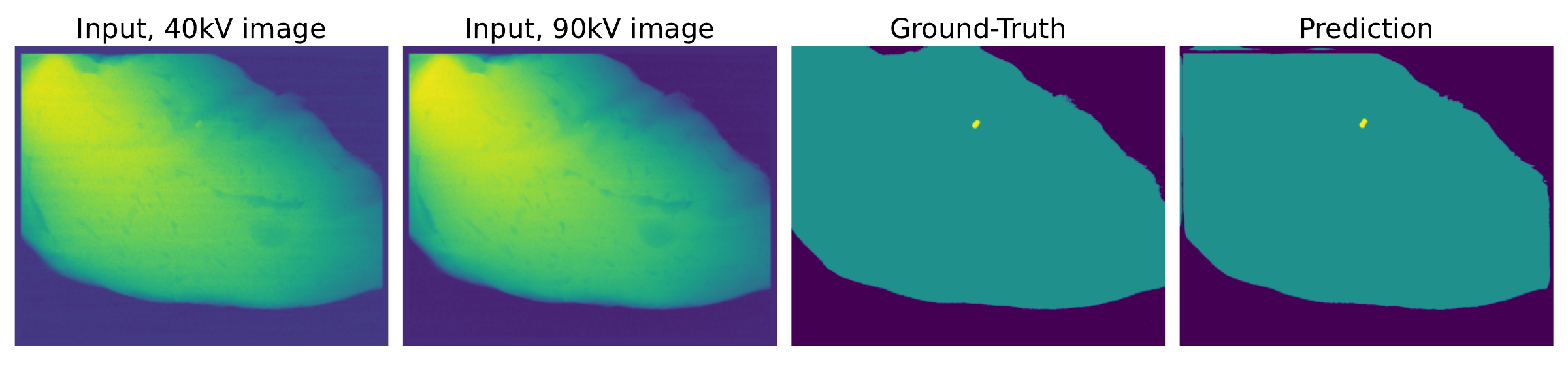}
\caption{Example of an accurate segmentation. The DCNN takes an image with two channels (40kV and 90kV) as input. The segmentation mask is compared to the ground-truth to evaluate accuracy.}
\label{fig:segm_example}
\end{figure}

When tested on the projections containing bone fragments, the network was segmenting foreign objects with a mean recall value of 75\%. An example of accurate segmentation is shown on Fig. \ref{fig:segm_example}. To assess the performance on boneless cases, we have converted the segmentation results into the detection results. Under this definition, the network scored 96\% accuracy. Thus, with long exposure time, the selected DCNN architecture was able to solve the FOD problem despite the complicated morphological structure of projections and the limited amount of training data.

\begin{figure}[ht]
\centering
\includegraphics[width=0.95\linewidth]{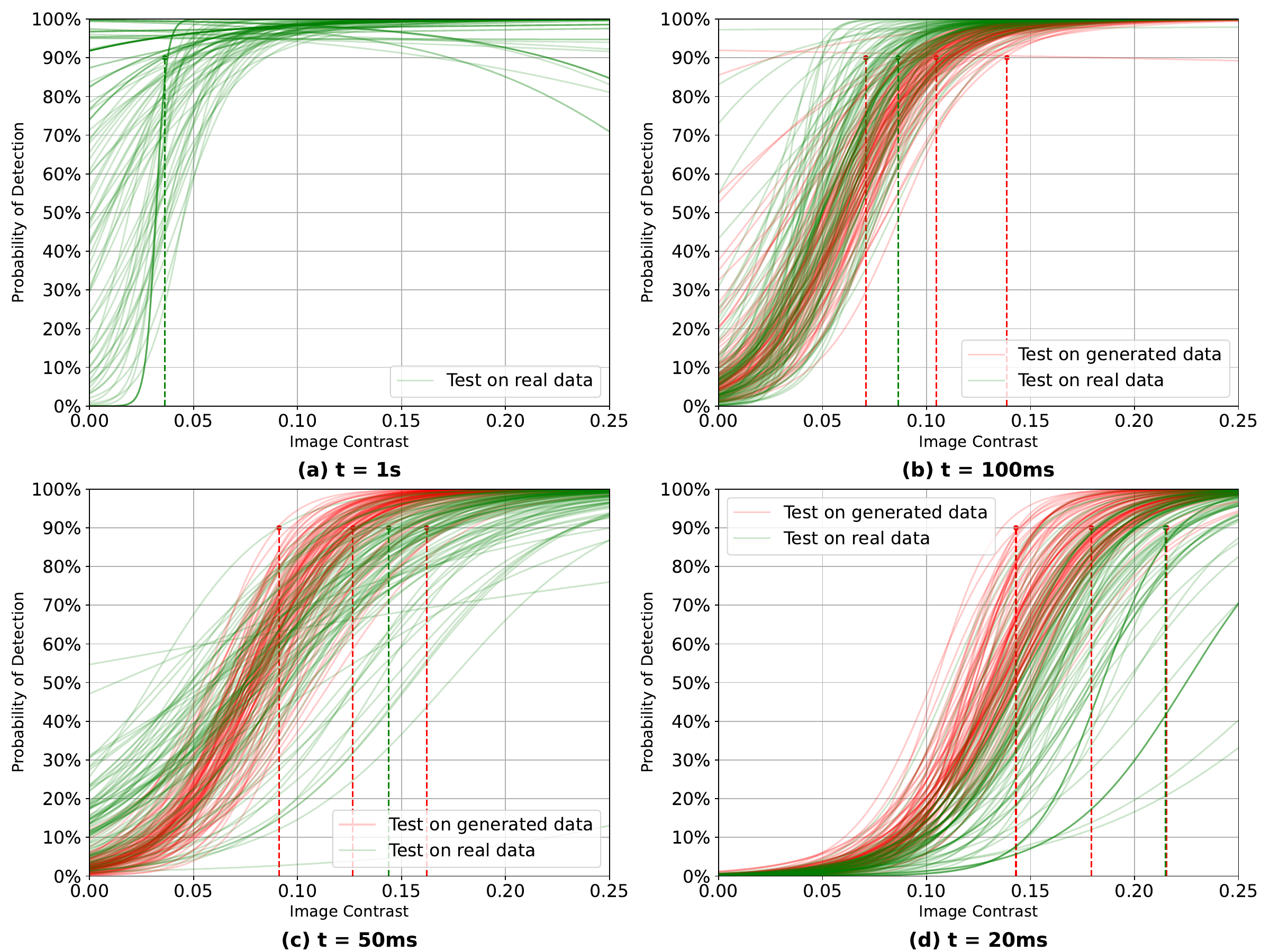}
\caption{POD curves representing the performance of the DCNNs trained on generated data for 4 different values of exposure time: (a) 1s, (b) 100ms, (c) 50ms, (d) 20ms. Each line corresponds to an instance of DCNNs, 100 instances were trained for every exposure time. Red lines correspond to performance on generated data, green lines - on real data. The value of image contrast corresponding to the detection probability $P = 90\%$ is similar for real and generated data, the difference lies withing the confidence interval.}
\label{fig:pod_curves}
\end{figure}

The generation approach and the setup parameters extracted from calibration were used to train DCNNs for FO segmentation with an exposure time of 100~ms, 50~ms, and 20~ms. We tested these networks on the second dataset containing small bone fragments in different locations. The POD curves as a function of FO contrast defined by \ref{eq:dr_def} are shown on Fig. \ref{fig:pod_curves}. We have observed that the value of contrast at which the probability of detection reaches 90\% is similar for tests on real and generated data taking into account the variance from deep learning training. These values are shown in Table \ref{tab:gen_real_comp}. Thus, the generation algorithms produced sufficiently accurate images, so the training on generated data was possible and the test results were similar for real and generated data.

\begin{table}[]
\begin{tabular}{|c|cc|}
\hline
\multirow{2}{*}{Exposure time, ms} & \multicolumn{2}{c|}{Contrast value for POD=90\%}          \\ \cline{2-3} 
                                   & \multicolumn{1}{c|}{Generated test data} & Real test data \\ \hline
1000                               & \multicolumn{1}{c|}{-}     & 0.03 $\pm$ 0.03\\ \hline
100                                & \multicolumn{1}{c|}{0.10 $\pm$ 0.02}     & 0.09 $\pm$ 0.02\\ \hline
50                                 & \multicolumn{1}{c|}{0.13 $\pm$ 0.02}     & 0.14 $\pm$ 0.17\\ \hline
20                                 & \multicolumn{1}{c|}{0.18 $\pm$ 0.02}     & 0.21 $\pm$ 0.04\\ \hline
\end{tabular}
\caption{Comparison of performance metrics for different values of exposure time with tests on real and generated data. Values of contrast corresponding to the high accuracy are similar between real and generated test data with respect to the uncertainty.}
\label{tab:gen_real_comp}
\end{table}

\begin{figure}[ht]
\centering
\includegraphics[width=0.95\linewidth]{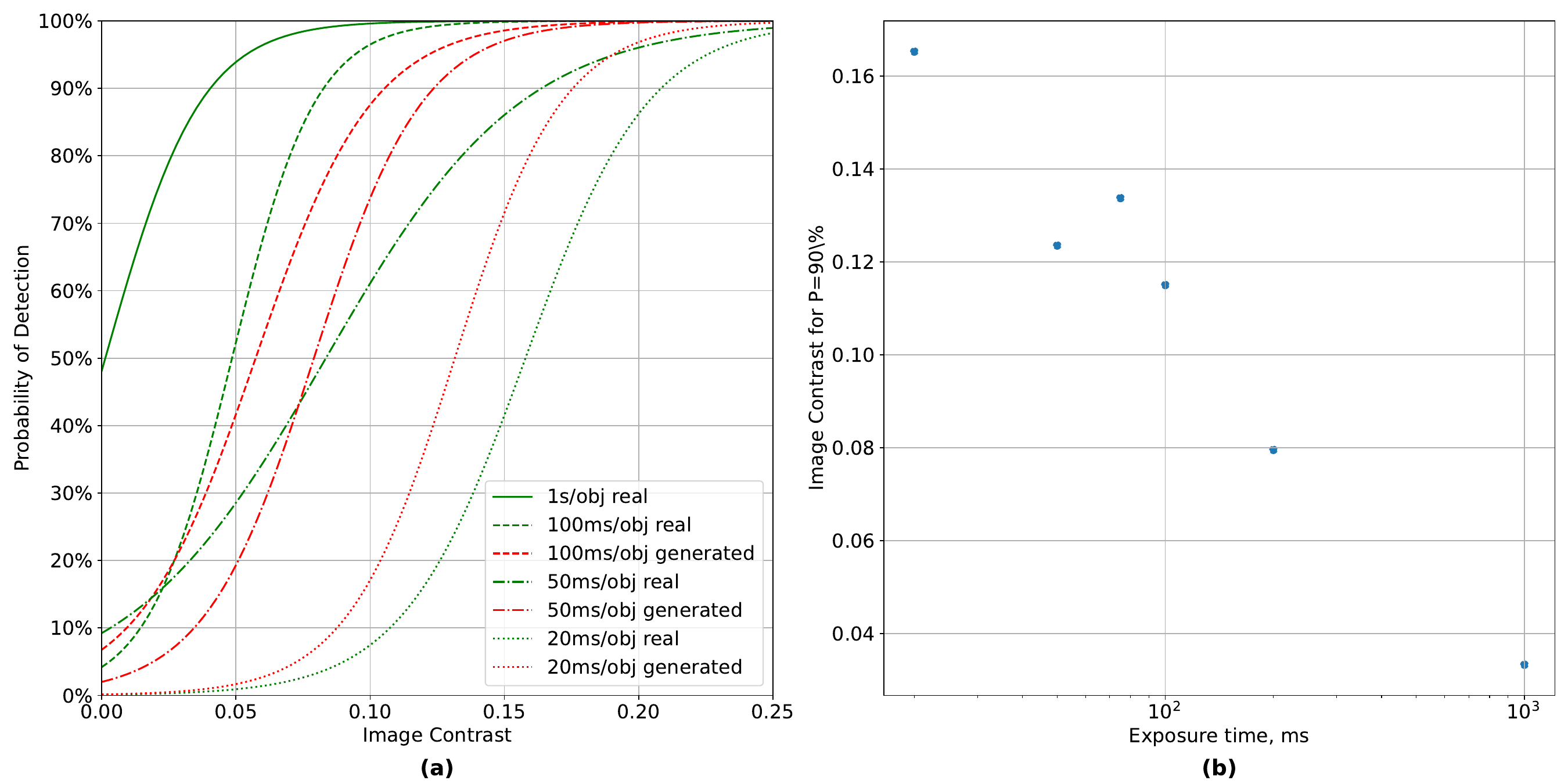}
\caption{(a) Comparison of POD curves corresponding to different values of exposure time. Each curve uses the average values of the fit parameters for curves shown on Fig. \ref{fig:pod_curves}. Lower exposure time moves the POD curve to the right, indicating a worse detection rate at the same values of contrast. (b) Image contrast is necessary for a good detection rate ($P = 90\%$) as a function of the exposure time (the plot uses the logarithmic scale for the exposure time).}
\label{fig:pod_same}
\end{figure}

Fig. \ref{fig:pod_same}a compares POD curves corresponding to different values of exposure time. Every curve corresponds to tests on generated data (except for 1~s where real data would be indistinguishable from generated), and the fit parameters are averaged over 100 iterations of DCNN training. The figure shows the expected effect of exposure time on detection: for the same value of contrast the performance gets worse with less exposure. Following the Methods section, this can be explained by a higher noise level. Using the image generator, Fig. \ref{fig:pod_same}a could be extended to other values of exposure time and used to find the optimal system design. Safety guidelines lead to a threshold for contrast that characterizes FOs that have to be detected consistently. POD curves show which exposure times lead to a sufficiently high detection probability for that contrast, and the smallest time could be chosen to balance high throughput and product safety. Assuming that the probability of detection $P=90\%$ is enough for reliable detection, we show on Fig. \ref{fig:pod_same}b how the necessary value of contrast changes with the exposure time. For Fig. \ref{fig:pod_same}b we have estimated the performance level for 200~ms and 75~ms, where no experimental data are available. We have also experimented with smaller values of exposure time between 20ms and 50ms, but a high variance of the performance estimate was observed. This might indicate that the accuracy of the generation method is not sufficient around 20ms.

\section*{Discussion}

Using image generation to predict inspection performance relies on the assumption that generated images could be similar enough to real ones. While we have checked this using POD curves, the problem of similarity between real and generated data could be addressed directly. Under low noise levels, even a difference between two images could be used to conclude if they are similar. However, high-throughput X-ray inspection is likely to work with noisy data. Thus, even two real images of the same object under the same conditions are never the same. To estimate similarity, it is necessary to check whether the noise pattern in two images corresponds to the same distribution. However, perfect correspondence is only expected for highly accurate Monte-Carlo algorithms which are not practical when the goal is to accelerate performance evaluation. Furthermore, it is unknown what level of similarity is necessary to get the same image analysis response. Thus, even if the accuracy of the generation model was quantitatively estimated, it would not be sufficient to conclude that it leads to the correct performance estimate.

We have chosen a data-driven definition of image similarity - matching the POD curves. While this approach does not characterize the accuracy of the generator in all possible problems under different system settings, there is a direct confirmation that the generated data lead to the same performance estimate as the real data in a particular problem of interest. We assume that interpolation in terms of system settings is possible. If the generator is accurate for a number of values of different settings, it could be used for the values in-between. Thus, with a small number of real datasets under different settings, a wider range could be explored with the image generator. 

The results show that a complete match between the POD curves for the real and generated data is unlikely. While the inaccuracy of the generator contributes to the difference between the POD curves, there is an inherent uncertainty in the proposed performance evaluation method. We indicate two main sources: the uncertainty of the DCNN performance and the uncertainty of the POD curves as a method of performance evaluation. To account for the random nature of DCNN training, we repeated the process 100 times and used the variance of the POD coefficients as a measure of the variance between the trained networks. Additional DCNN training was stopped when the standard deviation of the coefficients converged. Furthermore, the DCNN uncertainty is implicitly affected by biases in training and test data. If an image in the test dataset contains features not present during training (e.g. different shapes of bone or fillet), the network is more likely to fail. To ensure a good coverage of features, we used a large variety of bones and chicken fillets to make experimental datasets. Furthermore, the results on high-quality data showed that the network was able to perform the FOD with high accuracy, which would not be possible with significant biases in feature coverage.

The goal of the POD curve in our methodology is to provide a criterion (dual-energy quotient contrast $\Delta R$) that guarantees high detection accuracy. As POD curves are computed using a log-likelihood fit, the resulting criterion could only be computed accurately with a proper test dataset. It should fully capture the transition from POD $=$ 0\% to POD $=$ 100\% and have enough undetectable and always detectable cases. We observed that even a single outlier could significantly increase the uncertainty estimate for a POD curve. In the experimental data, it was crucial to have a wide range of $\Delta R$ in the test dataset, so even for $t = 20$~ms it was possible to achieve a high detection rate.

The main downside of the proposed methodology is that the high performance criterion is linked to the image property $\Delta R$. Industrial requirements would prefer to connect the performance to the physical properties of the foreign object, such as size, shape, and location in the main object. Since the POD curve is only a method of statistical analysis, it can be performed with respect to any feature. However, there is no guarantee that consistently high performance could be achieved with a simple constraint on a test sample. The image formation subsection shows how foreign object visibility depends on many parameters. The inspection system does not control the orientation of the FO with respect to the detector and the morphology of the main object around the FO. These issues are not unique to our methodology. The inability to control the view of the object affects all single projection inspection systems regardless of the image analysis and performance evaluation method. The second experimental dataset further highlights that foreign objects of the same size could be consistently detected or missed based on their location in the main object.

\section*{Conclusion}

We have presented a computationally efficient model of X-ray image generation. The proposed calibration procedure can be applied to a large variety of X-ray inspection setups and does not require specialized equipment. We have represented the performance of the X-ray inspection system under different experimental settings as POD curves. Our results showed that the image generation method leads to the same POD curves within the uncertainty limits. Thus, the generated data could be used to predict the performance of the system. Optimal acquisition parameters could be chosen based on the optimal POD curve. We demonstrated this approach on an example of bone fragment detection in chicken fillets with exposure time as a flexible experimental setting.

\section*{Acknowledgements}
We are grateful to TESCAN-XRE NV for their collaboration and support regarding the FleX-ray Laboratory.

\section*{Data availability}
Experimental datasets are available on Zenodo: \url{https://zenodo.org/doi/10.5281/zenodo.10579607}

\section*{Code availability}
Code for X-ray image generation, DCNN training and testing, and POD analysis can be found on Gihub: \url{https://github.com/vandriiashen/pod2settings}

\section*{Author contributions statement}

V.A., R.v.L., T.v.L., and K.J.B. conceived the research and the experiments, V.A. performed calibration of the experimental setup, implemented the image generation code, and acquired the experimental datasets, V.A., R.v.L., T.v.L., and K.J.B. analysed the results. V.A. wrote the manuscript and prepared the figures. R.v.L., T.v.L., and K.J.B. provided edits and comments to the manuscript. All authors reviewed the manuscript.

\section*{Statements and Declarations}
\subsection*{Competing interests}
The authors declare no competing interests.

\bibliography{main.bib}

\end{document}